\documentstyle[12pt,epsfig]{article}
\begin{document}

\begin{titlepage}
\rightline{March 2013}
\vskip 3cm
\centerline{\Large \bf
A dark matter scaling relation}
\vskip 0.3cm
\centerline{\Large \bf
 from mirror dark matter}

\vskip 1.8cm
\centerline{R. Foot\footnote{
E-mail address: rfoot@unimelb.edu.au}}

\vskip 0.7cm
\centerline{\it ARC Centre of Excellence for Particle Physics at the Terascale,}
\centerline{\it School of Physics, University of Melbourne,}
\centerline{\it Victoria 3010 Australia}
\vskip 3cm
\noindent
Mirror dark matter, and other similar dissipative dark matter
candidates, need an energy source to stabilize dark matter halos
around spiral galaxies. It has been suggested previously that ordinary
supernovae can potentially supply the required energy.
By matching the energy supplied to the halo from supernovae
to that lost due to radiative cooling,
we here derive a rough scaling relation, $R_{SN} \propto 
\rho_0 r_0^2$ ($R_{SN}$ is the supernova rate and $\rho_0, \ r_0$ the dark matter central
density and core radius).
Such a relation is consistent with dark matter properties inferred from studies of spiral galaxies with
halo masses larger than $3\times 10^{11} M_\odot$.
We speculate that other observed galaxy regularities might be explained within the framework of such
dissipative dark matter.

\end{titlepage}


A hidden sector exactly isomorphic to the ordinary
matter sector is required if one hypothesizes fundamental improper space-time symmetries which are unbroken \cite{flv}.
In such a theory, there is a mirror particle corresponding to every type of ordinary particle, except perhaps the graviton.
Thus, a spectrum of stable dark particles naturally arises. We denote these mirror particles with a prime ($'$),
$e', H', He', ...$. The symmetry implies that the masses of these particles are identical to their ordinary matter counterparts and that
the mirror particles interact with mirror gauge fields (such as the mirror photon) in a manner
completely analogous to the ordinary matter sector.

Mirror particles have emerged as an interesting candidate for
dark matter (for reviews and more
complete bibliography see e.g.\cite{review}).
Mirror dark matter can explain \cite{foot} the positive dark matter signals from the DAMA \cite{dama}, CoGeNT \cite{cogent}
and
CRESST-II \cite{cresst} direct detection experiments. This requires photon - mirror
photon kinetic mixing (defined below) of strength
$\epsilon \sim 10^{-9}$.
Mirror dark matter can also explain \cite{lss} the large-scale structure of the Universe (matter power spectrum and CMB)
in a similar way to
standard collisionless cold dark matter provided \cite{cmb} that $\epsilon \stackrel{<}{\sim} 3
\times 10^{-9}$.

On small scales mirror dark matter
has a number of distinctive features due to self interactions and dissipative interactions. In contrast to collisionless
particles,
galactic halos of spiral galaxies are composed predominately of mirror particles in a pressure supported spherical
plasma \cite{sph}.
There may also be a subcomponent
consisting of compact objects such as old mirror stars \cite{macho1} (gravitational lensing observations limit the
MACHO halo fraction of the Milky Way to be less than around 0.3 \cite{macho}).
Because mirror dark matter is dissipative,
an energy source is needed to stabilize the dark matter halos
around galaxies. [Without an energy source the mirror particles would collapse to a dark disk on a time
scale typically around a few hundred million years.] It has been speculated previously \cite{sph} that ordinary
supernovae can potentially supply the required energy if photon - mirror photon kinetic mixing exists \cite{he}:
\begin{eqnarray}
{\cal L}_{mix} = \frac{\epsilon}{2} F^{\mu \nu} F'_{\mu \nu}
\label{kine}
\end{eqnarray}
where $F_{\mu \nu}$
($F'_{\mu \nu}$) is the field strength tensor for the photon (mirror photon).
The physical effect of the kinetic mixing interaction is to induce a tiny ordinary
electric charge ($\propto \epsilon$) for the mirror charged particles \cite{holdom,flv}.
In the hot and dense core of type II supernovae mirror electrons and positrons can be produced from
plasmon decay processes \cite{raflelt}.
Thus ordinary supernovae can be a source of light mirror particles
as well as the ordinary neutrinos.
Indeed, it is estimated that around half of the core collapse supernova energy
is emitted by mirror particles
($e', \ \bar e', \ \gamma'$) if $\epsilon \sim 10^{-9}$ \cite{raflelt,sil}.
A significant fraction
of this energy might possibly be absorbed by the mirror particle halo.
We show here that a 
rough scaling relation, $R_{SN} \propto 
\rho_0 r_0^2$ ($R_{SN}$ is the supernova rate and $\rho_0, \ r_0$ the dark matter central density and core radius)
follows by
matching the energy supplied to the halo from supernovae
to that lost due to radiative cooling.
We find that this derived relation is
roughly consistent with dark matter properties inferred from studies
of spiral galaxies.
Although our discussion is in the context of mirror dark matter,
a similar conclusion could be obtained for
a class of dissipative dark matter candidates, 
such as those discussed recently in ref. \cite{china}. 

The physical picture then, is that galactic halos in spiral galaxies are composed predominately of mirror particles
$e', H', He',....$ in a self-interacting pressure supported halo.
Clue's about the chemical composition of this halo arise from early Universe cosmology.
Calculations indicate \cite{paolo1}
a primordial mirror helium mass fraction around 0.9 for $\epsilon \sim 10^{-9}$.
This suggests a halo composed primarily of $He'$ with perhaps a fraction of $H'$ and mirror metal components
(produced in mirror star formation at an earlier epoch). To a first approximation, we can consider the halo as composed
of
mirror helium, which for temperatures above around 40 eV should be fully ionized.
The radiative cooling rate, $\Gamma_{cool}$, of such a mirror particle plasma is given by the analogous expression
for ordinary matter plasma \cite{books}
\begin{eqnarray}
\Gamma_{cool} = \Lambda (T)\int n_{e'}^2 dV 
\end{eqnarray}
where $n_{e'}$ is the $e'$ number density, and $\Lambda (T)$ is the cooling function and has units of
${\rm erg} \ {\rm cm}^3 \ {\rm s}^{-1}$.
For a fully ionized mirror helium plasma, $n_{e'} \simeq 2\rho_{dm}/m_{He}$.

Rotation curves in spiral galaxies are
well fit \cite{paolo9} with a baryonic component described by a Freeman disk \cite{freeman} and
a spherically distributed cored dark matter component with the Burkert profile \cite{bp}:
\begin{eqnarray}
\rho_{dm} = {\rho_0 r_0^3 
\over (r + r_0)(r^2 + r_0^2)
}
\end{eqnarray}
where $r_0, \ \rho_0$ is the dark matter core radius and central density respectively.
However, as discussed in ref. \cite{sph},
a spherical self gravitating isothermal
gas of particles requires $\rho \propto 1/r^2$. If the temperature is not isothermal but
rises in the inner region then it might be possible
to produce a cored distribution.
This motivates the
quasi-isothermal profile with $\rho_{dm} = {\rho_0 r_0^2 \over (r^2 + r_0^2)}$.
Note that if, instead of the Burkert profile we adopted
the quasi-isothermal profile, which also provides a reasonable fit to the
rotation curves of spiral galaxies, this would not significantly affect
our subsequent analysis.
Anyway, assuming a mirror dark matter halo composed (predominately) of an ionized plasma with the Burkert
density profile, it follows that
\begin{eqnarray}
\Gamma_{cool} = \Lambda (T) \rho_0^2 r_0^3 (4/m_{He}^2) I
\end{eqnarray}
where
\begin{eqnarray}
I \equiv 4\pi \int_0^\infty {x^2 \over (1+x)^2(1 +x^2)^2} \ dx 
\simeq 1.34
\ .
\end{eqnarray}
Thus, we find:
\begin{eqnarray}
\Gamma_{cool} \simeq \left({\Lambda (T) \over 10^{-23} \ {\rm erg \ cm}^3/{\rm s}} \right) 
\left( {\rho_0 r_0 \over 10^{2.2} M_\odot/{\rm pc}^2}\right)^2
\left( {r_0 \over 10\ {\rm kpc}}\right) \ 4\times 10^{43} \ {\rm erg/s}
\ .
\end{eqnarray}
On the other hand, the rate at which the halo absorbs the energy from supernovae is:
\begin{eqnarray}
\Gamma_{SN} &=& f_{SN} \langle E_{SN} \rangle R_{SN}
\nonumber \\
& \simeq & 
\left( {f_{SN} \over 0.1}\right)
\left({\langle E_{SN} \rangle \over 3\times 10^{53}\ {\rm erg}} \right) \left({R_{SN} \over 0.03 \ {\rm yr}^{-1}} \right) 
\ 3\times 10^{43} \ {\rm erg/s}
\label{fsn}
\end{eqnarray}
where $f_{SN}$ is the proportion of the supernova total energy, $E_{SN}$, absorbed by the halo and $R_{SN}$
is the galactic supernova rate.
Equating $\Gamma_{cool}$ with $\Gamma_{SN}$ for the Milky Way galaxy
implies $f_{SN} \sim 0.1$ with an order of magnitude uncertainty.

In this picture, $\Gamma_{cool} \simeq \Gamma_{SN}$ should hold for any galaxy, not just the Milky Way.
Imposing this condition, and using the expected scaling $f_{SN} \propto \rho_0 r_0$, (which assumes an optically thin halo)
suggests a scaling relation:
\begin{eqnarray}
R_{SN} \propto \Lambda (T) \rho_0 r_0^2
\ .
\end{eqnarray}
The idea is that the dynamics can keep this relation satisfied. If $\Gamma_{SN} > \Gamma_{cool}$
($\Gamma_{SN} < \Gamma_{cool}$) then the
halo should expand (contract), thereby decreasing (increasing) the star formation rate, and hence $R_{SN}$,
until $\Gamma_{SN} \approx \Gamma_{cool}$.

What is the temperature $T$? For an isothermal halo in hydrostatic equilibrium, we expect \cite{sph}
\begin{eqnarray}
T \approx {1 \over 2} \bar m v_{rot}^2
\label{T}
\end{eqnarray}
where $\bar m$ is the mean mass of the particles in the halo ($\bar m \approx 1.3$ GeV for a mirror helium dominated halo)
and $v_{rot}$ is the asymptotic value of the rotational velocity.
For spiral galaxies, halo masses have values
$3 \times 10^{10} \ M_\odot 
\stackrel{<}{\sim} M_h \stackrel{<}{\sim} 3\times 10^{13} \ M_\odot$.
For such halo masses, $v_{rot}$ has the range $50 \ {\rm km/s} \stackrel{<}{\sim} v_{rot} \stackrel{<}{\sim} 500$ km/s,
and Eq.(\ref{T}) suggests a rough temperature range:
\begin{eqnarray}
10 \ {\rm eV} \stackrel{<}{\sim} T \stackrel{<}{\sim} 1000 
\ {\rm eV}\ .
\label{range}
\end{eqnarray}
At the lowest temperature's mirror helium will not be fully ionized and the cooling rate can become dramatically
suppressed. 
This might explain why low mass halos $< 10^{10}\ M_\odot$ hosting disk systems are not detected.
For temperatures above $\sim 20$ eV, mirror helium is fully ionized.
 and for $40 \ {\rm eV} \stackrel{<}{\sim} T 
\stackrel{<}{\sim} 1000 \ {\rm eV}$, $\Lambda (T)$ typically varies by
only a factor of 2-3 \cite{books}. [This is relatively minor compared to the variation of, say, $r_0$ over this range of
halo masses,
which is around 2 orders of magnitude.]
Taking the rough approximation of $\Lambda (T)$ as constant in this temperature range,
yields the scaling relation:
\begin{eqnarray}
R_{SN} \propto \rho_0 r_0^2 \ .
\label{p}
\end{eqnarray}

Is the above relation satisfied by spiral galaxies? A set of scaling relations have been derived from
observations of spiral
galaxies \cite{kf,dsds,sal2,sal3}
and summarized recently in ref. \cite{sal4}.
These relate $\rho_0, r_0, M_h$ and r*-band luminosity, $L_r$:
\newpage 
\begin{eqnarray}
log\left( {\rho_0 r_0 \over M_\odot {\rm pc}^{-2}} \right) &\simeq  & 2.2 \pm 0.25 \nonumber \\
{L_r \over 1.2\times 10^{10} L_\odot} &\simeq & 
{\left({M_h \over 3 \times 10^{11} M_\odot}\right)^{2.65} \over 
1 + \left( {M_h \over 3\times 10^{11} M_\odot}\right)^{2.00} 
}
\nonumber \\
log \left( {r_0 \over {\rm kpc}} \right) &\simeq & 0.66 + 0.58 \ log \left( {M_h \over 10^{11} M_\odot}  
\right)
\ .
\label{rel1}
\end{eqnarray}
These relations imply that for spiral galaxies with $M_h \stackrel{>}{\sim}  3 \times 10^{11} M_\odot$,
\begin{eqnarray}
L_r  \propto r_0^{1.1}
\ .
\label{fe}
\end{eqnarray}
Observations\cite{sn} indicate that the galactic B-band luminosity, $L_B$, scales wtih the type II supernova rate
as $L_B \propto R_{SN}^{1.3}$,
with an uncertainty in the exponent around 0.1.
Neglecting the expected minor difference
between the scaling of $L_B$ and $L_r$
we arrive at a
rough `empirical' scaling relation:
\begin{eqnarray}
R_{SN} \propto r_0^{0.8}
\ .
\end{eqnarray}
This relation is consistent, within the uncertainties, to the rough `theoretical' scaling relation arrived at in
Eq.(\ref{p}) (given that
$\rho_0 r_0$
is observed to be approximately constant).
This provides some observational support to the notion that ordinary supernova supply the energy
needed to stabilize halos in spiral galaxies, at least for
$M_h \stackrel{>}{\sim} 3 \times 10^{11} M_\odot$.

The above analysis has assumed that a significant fraction of supernova energy can be
transmitted to the halo. 
How reasonable is this assumption?
Let us assume a kinetic mixing parameter $\epsilon \sim 10^{-9}$, so that around half of type II supernova
energy is converted into $e', \ \bar e'$ emitted from the core with energies $\sim $ MeV.
One could imagine that the huge number of MeV $e', \bar e'$ injected into a volume [($\sim 1\ {\rm pc})^3$]
around ordinary supernova will radiatively cool, converting most of their energy into mirror
photons.
The energy spectrum of these mirror photons is of course very hard to predict
but it might be have some vaguely similar features to the 
$\gamma$ spectrum of ordinary Gamma Ray Bursts (GRB's). GRB's
feature a wide spectrum of energies with mean around 700 keV with a few percent of energy radiated below 10 keV.
In any case,
these mirror photons will then heat the mirror particle halo, potentially supplying the energy lost from
the halo due to radiative cooling.
Whether this can happen depends on how strongly the mirror photons scatter off mirror electrons, both bound and free.

Consider first
the scattering off free mirror electrons, i.e. elastic (Thomson) scattering with
$E'_\gamma$ independent cross-section $\sigma_T = 6.7\times 10^{-25}\ {\rm cm^2}$.
We
estimate that the optical depth due to elastic scattering for $\gamma'$  propagating out
from the galactic center is
\begin{eqnarray}
\tau_{ES} &= & \int_0^{\infty} \sigma_T n_{e'} dr 
 \ \approx   \ 0.78 \sigma_T \rho_0 r_0 \left( {2 \over m_{He}}\right) \sim 0.006
\end{eqnarray}
where we obtained $\rho_0 r_0$ from Eq.(\ref{rel1}).
We expect $f_{SN} {\sim}  \tau$ and
Eq.(\ref{fsn}) then suggests that elastic scattering is probably not frequent enough to supply enough heat to the halo to
stabilize it.
The cross-section is at least an order of magnitude too small.
However if the halo contains a significant proportion of heavy mirror elements -  necessary to explain the direct
detection
experiments \cite{foot} -
then the photoelectric cross-section of heavy mirror elements
can easily dominate over the elastic cross-section for a large range of energies. This was noted in ref. \cite{sph} and
we expand upon this point here.

Heavy elements, such as $A' = O',Si',Fe'$, are not completely ionized but have their atomic inner shells filled.
The total photoelectric cross-section (in units with $\hbar = c = 1$) for the inner K shell mirror electrons of a mirror
element with atomic number, $Z$, is given approximately by \cite{book5}
\begin{eqnarray}
\sigma_{A'} (E'_\gamma) = {16\sqrt{2} \pi \over 3m_e^2 } \alpha^6  Z^5 \left[ {m_e \over E'_{\gamma}
}\right]^{7/2} 
\ .
\label{16}
\end{eqnarray}
Evidently, the photoelectric cross-section
decreases with mirror photon energy like $(E'_\gamma)^{-7/2}$ and, of course,
$E'_\gamma$ must be larger than the mirror electron binding energy of the particular element concerned.
The contribution to the optical depth
due to such inelastic scattering for $\gamma'$ propagating out from the galactic center is
\begin{eqnarray}
\tau_{IS} &= & \sum_{A'} 2\int_0^{\infty} \sigma_{A'} n_{A'} dr 
\nonumber \\
&\sim & \sum_{A'} 2\rho_0 r_0 \sigma_{A'} \left[ {\xi_{A'} \over m_{A'}}
\right]
\end{eqnarray}
where $\xi_{A'}$ is the proportion by mass of the mirror metal component, $A'$ (e.g. $A' = O', Si', Fe',...$) and
we have included a factor of two since there are two K shell mirror electrons.
For illustrative purposes we have
evaluated the total optical depth, including both elastic and inelastic scattering (the latter assumed dominated by K
shell bound
electron scattering
as discussed above) for an example with a $2\%$ metal component with $\xi_{C'} = \xi_{O'} = \xi_{Si'} = \xi_{Fe'} =
0.005$.
The result is shown in figure 1. This figure indicates that for mirror photon energies
\begin{eqnarray}
0.4\ {\rm keV} \stackrel{<}{\sim}
E'_\gamma \stackrel{<}{\sim} 30 \ {\rm keV}
\label{ran}
\end{eqnarray}
inelastic scattering of mirror photons can dominate over elastic scattering.
Thus, it might actually be possible for supernovae to transfer a significant part of their
energy to the halo in a fairly efficient manner \footnote{
Observe that the energy is transmitted initially to the mirror electron component rather than the mirror nuclei.
The liberated $e'$ will interact with the plasma primarily heating the $e'$ component.  The cooling processes also
primarily cool the $e'$ component rather than the mirror nuclei. Thus, to a first approximation it appears reasonable to assume
that the plasma is locally described by a single temperature, $T$. Of course $T$ can have some radial
dependence, but such details are beyond the scope of this rough analytic analysis.
}.

\centerline{\epsfig{file=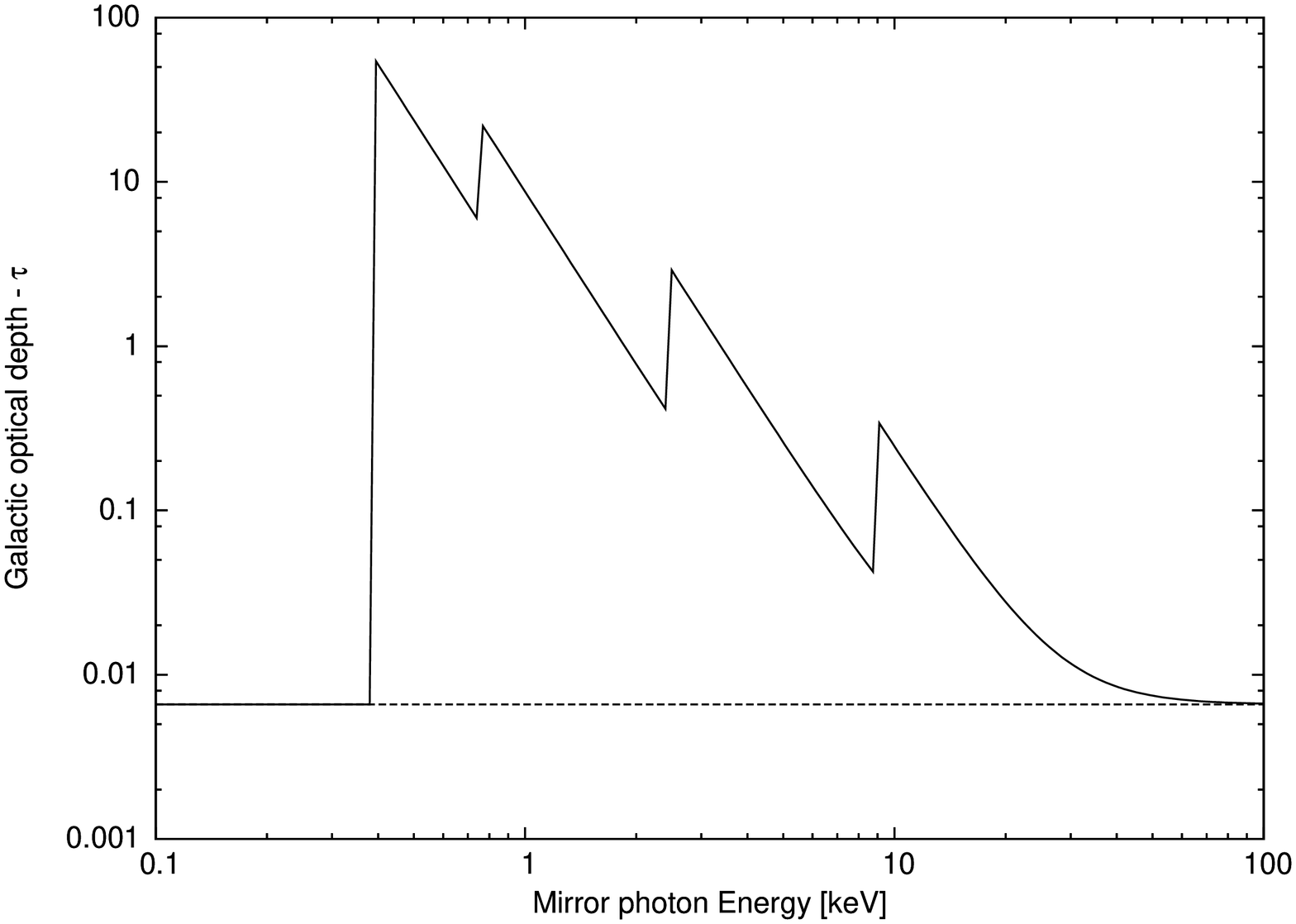, angle=270, width=12.8cm}}
\vskip 0.5cm
\noindent
{\small Figure 1: Galactic optical depth, $\tau$, versus mirror photon energy.
Also shown is the contribution to the optical depth due to elastic (Thomson) scattering, $\tau_{ES}$ (dotted line).  }
\vskip 1.0cm

If the halos of spiral galaxies are supported by the energy from supernovae then
it should be possible to make much more progress in understanding the
structural properties of spiral galaxies.  Additional scaling relations, possible existence of a core, etc., are
anticipated if all the relevant astrophysics is understood. 
It is conceivable, for example,
that the scaling relation, $\rho_0 r_0 \propto constant$ might be related to the
details of energy transport since the optical depth is also proportional to $\rho_0 r_0$.
One might suspect that the core region arises, at least in part, due to the heating up of 
the inner region of the halo due to the opacity.
Below we give some further thoughts on the subject.

The scaling relation, Eq.(\ref{p}), was derived from only 
the very general condition that the total energy lost due to radiative cooling
is replaced by the total energy input.
Of course in any particular volume element the energy input must match the energy output 
for a steady state configuration.
Let us first consider a toy model, where we model the baryonic component as a point source
whose energy output supports a spherical mirror dark matter halo. 
That is, we can assume a mirror photon luminosity, $L$, originating at $r=0$.
The energy going into a volume element, $dV = 4\pi r^2 dr$, assuming mirror radiation dominates the energy
transport is
\begin{eqnarray}
d{\cal E}_{in} &=& F \ \sum_{A'} \sigma_{A'} n_{A'} dV  
\nonumber \\
 &=&  {L \over 4\pi r^2} \sum_{A'} \sigma_{A'} n_{A'} dV  
\end{eqnarray}
where $\sigma_{A'}$ is 
the photoelectric cross-section given in Eq.(\ref{16}) and $n_{A'}$ is the $A'$ metal number density $A' = O', Si', Fe',...$.
We have assumed in this toy model that
the optical depth, $\tau \equiv \int \sum \sigma_{A'} n_{A'} dr \ll 1$ so that the $\gamma'$ flux, $F$,  scales as $\sim 1/r^2$.
The energy going out of the same volume element is
\begin{eqnarray}
d{\cal E}_{out} = \Lambda (T) n_{e'}^2 dV
\ .
\end{eqnarray}
Matching $d{\cal E}_{in} = d{\cal E}_{out}$ implies
\begin{eqnarray}
n_{e'} = F \ {\sum_{A'} \sigma_{A'} n_{A'} \over \Lambda(T) n_{e'}} = {L \over 4\pi r^2} {\sum \sigma_{A'} n_{A'}
\over \Lambda(T) n_{e'} }
\ .
\end{eqnarray}
The ratio $n_{A'}/n_{e'}$ is expected to be roughly independent of $r$ (given the charge neutrality of the plasma).
It follows that $n_{e'} \propto 1/r^2$, and recall that this behaviour also matches the condition from hydrostatic equilibrium
of a pressure supported self gravitating spherical distribution with a common (i.e. independent of $r$) temperature,
$T$ (see e.g. \cite{sph}).
Thus a self consistent `toy model' emerges, except it is unphysical at $r = 0$.


Let us now perturb this picture, by modelling the energy source, not as a point, but as a distribution extended
over a distance $\sim$ $r_D$. In this more realistic case we expect $n \sim 1/r^2$ for $r \gg r_D$ and a softer behaviour 
for $r \stackrel{<}{\sim} few\ r_D$.
Specifically, consider supernova sources  distributed in a Freeman disk with surface density, 
\begin{eqnarray}
\Sigma (\stackrel{\sim}{r})= {M_D \over 2\pi r_D^2} e^{-\stackrel{\sim}{r}/r_D} \ . 
\end{eqnarray}
It is convenient to use cylindrical co-ordinates 
$(\stackrel{\sim}{r}, \stackrel{\sim}{\theta}, \stackrel{\sim}{z})$ with the disk at $\stackrel{\sim}{z}=0$.
The average flux at the point 
$P = (r_1, 0, z_1)$ is then
\begin{eqnarray}
F(r_1,z_1) = {L \over 4\pi M_D} \int \int 
{\Sigma (\stackrel{\sim}{r}) \over 
\stackrel{\sim}{r}^2 - 2\stackrel{\sim}{r} r_1 \cos \stackrel{\sim}{\theta} + r_1^2 + z_1^2 } \ 
\stackrel{\sim}{r} d\stackrel{\sim}{r} d\stackrel{\sim}{\theta}
\ .
\end{eqnarray}
In this case,  
matching $d{\cal E}_{in} = d{\cal E}_{out}$ implies
\begin{eqnarray}
n_{e'} = {F(r_1, z_1) \over \Lambda (T)} \ 
{\sum_{A'} \sigma_{A'} n_{A'} \over n_{e'}}
\ .
\end{eqnarray}
One can indeed show that $F(r_1,z_1) \propto 1/r^2$ (where $r^2 = r_1^2 + z_1^2$) for $r \gg r_D$ and has 
a much softer behaviour for $r \stackrel{<}{\sim} few \ r_D$ with  $F(r_1,z_1) \sim log(r)$  as $r \to 0$. 
This suggests a rough scaling behaviour of
$r_0 \propto r_D$, for which there is some evidence \cite{dsds}.
This is all very interesting, however 
more detailed studies are clearly needed to rigorously check these ideas.


In conclusion, we have considered galaxy structure within mirror dark matter - a dissipative and self-interacting 
dark matter candidate.
For this type of dark matter,
an energy source is needed to stabilize dark matter halos in spiral galaxies such as the Milky Way.
Previously \cite{sph} it has been speculated that ordinary supernovae can supply the required energy if photon-mirror photon
kinetic mixing of strength $\epsilon \sim 10^{-9}$ exists.
We have shown here that this argument motivates a
rough scaling relation, $R_{SN} \propto \rho_0 r_0^2$
($R_{SN}$ is the supernova rate and $\rho_0, \ r_0$ the dark matter central density and core radius).
Interestingly, this scaling relation
is consistent with the dark matter properties inferred from recent studies
of spiral galaxies.
We have also presented some speculative reasoning suggesting that more
detailed studies with this type of dark matter candidate might lead to much more progress
in understanding the structure of galaxies.

\vskip 0.5cm
\noindent
{\bf Note added.} Following this work, more detailed numerical analysis of the problem
of galaxy structure within the mirror dark matter framework have been undertaken in \cite{footmd1,footmd2}.

\vskip 0.9cm
\noindent
{\large \bf Acknowledgments}

\vskip 0.2cm
\noindent
The author would like to thank
Shunsaku Horiuchi and Paolo Salucci for useful discussions/correspondence.
This work was supported by the Australian Research Council.

\end{document}